\documentclass{bmcart}

\usepackage[utf8]{inputenc} 

\usepackage{url,comment}
\usepackage{graphicx}

\startlocaldefs
\endlocaldefs

\begin{document}

\begin{frontmatter}

\begin{fmbox}
\dochead{Research}


\title{Online Abuse toward Candidates during the UK General Election 2019: Working Paper}


\author[
   addressref={aff1},                   
   corref={aff1},                       
   email={g.gorrell@sheffield.ac.uk}   
]{\inits{GM}\fnm{Genevieve} \snm{Gorrell}}
\author[
   addressref={aff1},
   email={m.e.bakir@sheffield.ac.uk}
]{\inits{ME}\fnm{Mehmet E} \snm{Bakir}}
\author[
   addressref={aff1},
   email={i.roberts@sheffield.ac.uk}
]{\inits{IR}\fnm{Ian} \snm{Roberts}}
\author[
   addressref={aff1},
   email={m.a.greenwood@sheffield.ac.uk}
]{\inits{MA}\fnm{Mark A} \snm{Greenwood}}
\author[
   addressref={aff1},
   email={k.bontcheva@sheffield.ac.uk}
]{\inits{KL}\fnm{Kalina} \snm{Bontcheva}}



\address[id=aff1]{
  \orgname{Department of Computer Science, Sheffield University}, 
  \street{Regent Court, 211 Portobello},                     %
  \city{Sheffield},                              
  \cny{UK}                                    
}



\end{fmbox}


\begin{abstractbox}

\begin{abstract} 
The 2019 UK general election took place against a background of rising online hostility levels toward politicians and concerns about polarisation. We collected 4.2 million tweets sent to or from election candidates in the six week period spanning from the start of November until shortly after the December 12th election. We found abuse in 4.46\% of replies received by candidates, up from 3.27\% in the matching period for the 2017 UK general election. Abuse levels have also been climbing month on month throughout 2019. The topic of UK withdrawal from the European Union drew an elevated level of abuse for candidates regardless of political party. Attention (and abuse) focused mainly on a small number of high profile politicians. Abuse is ``spiky'', triggered by external events such as debates, or controversial tweets. Abuse escalated throughout the campaign period. Men received more abuse overall; women received more sexist abuse. MPs choosing not to stand again had received more abuse during 2019.

\end{abstract}


\begin{keyword}
\kwd{UK general election}
\kwd{online abuse}
\kwd{twitter}
\kwd{politics}
\end{keyword}


\end{abstractbox}
%

\end{frontmatter}



\section*{Introduction}

Awareness is increasing of the social changes brought about through the rise of social media, and the consequences this can have politically. One area of concern is intimidation in public life. Where individuals are intimidated from participating in politics, democracy can be compromised. Dialogue on the subject takes the form of media\footnote{\small{\url{www.bbc.co.uk/news/election-2019-50687425}}} and government\footnote{\small{\url{https://www.gov.uk/government/consultations/online-harms-white-paper}}} engagement, research effort and innovation from the platforms.\footnote{\small{\url{https://blog.twitter.com/en_us/topics/product/2018/Serving_Healthy_Conversation.html}}} Yet there is much work still to be done in understanding the causes of internet toxicity and in forming an effective response.

In 2016, the UK voted to withdraw from the European Union in a close referendum that left parliament, as well as the nation, divided. Two general elections have followed, as successive prime ministers have sought to strengthen their majority in order to empower themselves in the withdrawal negotiations. In the context of heightened national feelings regarding ``Brexit'', it will be no surprise to many that online abuse toward politicians in the UK has increased. The 2019 UK general election offers an opportunity to explore in depth the factors that went into creating the experience political candidates had of being abused online.

Several areas are of particular interest. Using natural language processing we identify abuse and type it according to whether it is political, sexist or simply generic abuse. This enables a large-scale quantitative investigation. We present findings about what abuse is being sent and who it is being sent to. We show how the quantity of abuse being sent to politicians varies across the campaign period, across the year and in comparison with the general election in 2017. We compare the experience of MPs who chose to stand down vs. those who chose to stand again. We explore how the topic under discussion affects abusive responses, and how that varies between parties. We also investigate how abusive responses to individuals played out in the context of the key events of the campaign period. The work is of interest not only as a detailed data source regarding Twitter abuse toward politicians during the UK 2019 general election, but also in the context of the body of work tracking online abuse in political contexts.

\paragraph{Warning} In describing our work, we make frequent use of strong, offensive language, including slurs against minority and marginalised groups. This may be distressing for some readers.

\section*{Related Work}

As the effect of abuse and incivility in online political discussion has come to the fore in public discussion, the subject has begun to be seriously investigated researchers~\cite{rheault2019politicians,theocharis2016bad}. Binns and Bateman~\cite{binns2018and} review Twitter abuse towards UK MPs in early 2017. Gorrell \textit{et al}~\cite{gorrell2018twits} compare similar data from both the 2015 and 2017 UK general elections. Ward \textit{et al}~\cite{ward2017turds} explore a two and a half month period running from late 2016 to early 2017. Greenwood \textit{et al}~\cite{greenwood2019online} extends work presented by Gorrell \textit{et al}~\cite{gorrell2018twits} to span four years.

Women and ethnic minority MPs say that they receive worrying abuse~\cite{akhtar2019prevalence}, and abuse toward women has emerged as a topic of particular concern~\cite{rheault2019politicians,southern2019twitter,stambolieva2017methodology,delisle2019large}. Pew~\cite{pew2017} find that women are twice as likely as men to receive sexist abuse online, and are also more likely to perceive online abuse as a serious problem. Gorrell \textit{et al}~\cite{gorrell2019race} present findings for the first three quarters of 2019, with an emphasis on racial and religious tensions in UK politics. They find ethnic minority MPs, in addition to receiving more racist abuse, also receive more sexist abuse, and that women receive more sexist abuse. Rheault \textit{et al}~\cite{rheault2019politicians} find incivility toward women politicians increases with visibility, which they suggest relates to the extent of gender norm violations. Broadly speaking, the emerging picture is one in which women in politics are generally treated somewhat more politely than men, but within that, subjected to a lesser but more sinister volume of misogyny specific to them.

In addition to the implications for representation, as marginalised groups may feel pushed, or scared, out of the field, concern is increasing about offline threats to public figures also. Gorrell \textit{et al}~\cite{gorrell2019race} follow up on findings presented in the context of a BBC investigation\footnote{\small{\url{https://www.bbc.co.uk/news/uk-politics-49247808}}} into the increasing levels of threat and danger MPs are exposed to.

The quantitative work presented here depends on automatic detection of abuse in large volumes of Twitter data. A significant amount of work exists on the topic of automatic abuse detection within the field of natural language processing, often in the context of support for platform moderation. Schmidt and Wiegand~\cite{schmidt2017survey} provide a review of prior work and methods, as do Fortuna and Nunes~\cite{fortuna2018survey}. Whilst unintended bias has been the subject of much research in recent years with regards to making predictive systems that do not penalize minorities or perpetuate stereotypes, it has only just begun to be taken up within abuse classification \cite{park2018reducing}; unintended bias, such as an increased false positive rate for certain demographics, is a serious issue for sociological work such as ours. For that reason and others we adopt a rule-based approach here, as discussed below. More broadly, a biased dataset is one in which it is possible to learn classifications based on features that are unconnected to the actual task. Wiegand \textit{et al}~\cite{wiegand2019detection} share performance results for several well known abuse detection approaches when tested across domains.

\section*{Corpus and Methods}

Our work investigates a large tweet collection on which a natural language processing has been performed in order to identify abusive language, the politicians it is targeted at and the topics in the politician's original tweet that tend to trigger abusive replies, thus enabling large scale quantitative analysis. It includes, among other things, a component for MP and candidate recognition, which detects mentions of MPs. Topic detection finds mentions in the text of political topics (e.g. environment, immigration) and subtopics (e.g. fossil fuels). The list of topics was derived from the set of topics used to categorise documents on the gov.uk website\footnote{e.g. \small{\url{https://www.gov.uk/government/policies}}}, first seeded manually and then extended semi-automatically to include related terms and morphological variants using TermRaider\footnote{\small{\url{https://gate.ac.uk/projects/arcomem/TermRaider.html}}}, resulting in a total of 1046 terms across 44 topics.
This methodology is presented in more detail by~\cite{greenwood2019online}, with supporting materials also available online.\footnote{\label{note1} \small{\url{https://gate-socmedia.group.shef.ac.uk/election-analysis-and-hate-speech/ge2019-supp-mat/}}} However abuse detection has been extended since previous work, and is therefore explained in the next section.

\subsection*{Identifying Abusive Texts}

A rule-based approach was used to detect abusive language. An extensive vocabulary list of slurs, offensive words and potentially sensitive identity markers forms the basis of the approach.\footnote{\label{note1} \small{\url{https://gate-socmedia.group.shef.ac.uk/election-analysis-and-hate-speech/ge2019-supp-mat/}}} The slur list contained 1081 abusive terms or short phrases in British and American English, comprising mostly an extensive collection of insults, racist and homophobic slurs, as well as terms that denigrate a person's appearance or intelligence, gathered from sources that include \url{http://hatebase.org} and Farrell \textit{et al} \cite{farrell2019exploring}.

Offensive words such as the ``F'' word don't in and of themselves constitute abuse, but worsen abuse when found in conjunction with a slur, and become abusive when used with an identity term such as ``black'', ``Muslim'' or ``lesbian''. Furthermore, a sequence of these offensive words in practice is abusive. 131 such words were used; examples include ``f**king'', ``sh*t'' and ``fat''. Similarly, identity words aren't abusive in and of themselves, but when used with a slur or offensive word, their presence allows us to type the abuse. 451 such words were used.

Making the approach more precise as to target (whether the abuse is aimed at the politician being replied to or some third party) was achieved by rules based on pronoun co-occurrence. In the best case, a tight pronoun phrase such as ``you idiot'' or ``idiot like her'' is found, that can reliably be used to identify whether the target is the recipient of the tweet or a third party. Longer range pronoun phrases are less reliable but still useful. However, large numbers of insults contain no such qualification and are targeted at the tweet recipient, such as for example, simply, ``Idiot!''. Unless these are plurals, we count these. The approach is generally successful, but due to a high incidence of people making racist remarks to politicians about third parties, it is less successful for racism and religious intolerance, meaning that a politician to whom people make racist remarks may appear as having received racist abuse.

Data from Kaggle's 2012 challenge, ``Detecting Insults in Social
Commentary''\footnote{\small{\url{https://www.kaggle.com/c/detecting-insults-in-social-commentary/data}}}, was used to evaluate the success of the approach. The training set was used to tune the terms included. On the test set, our approach was shown to have an accuracy of 80\%, and a precision/recall/F1 of 0.72/0.47/0.57. This precision is considered sufficient for empirical work (being greater than 0.7~\cite{rijsbergen1979v}). However there is a long tail of linguistically more complex abuse that is hard to identify with sufficient precision, and therefore recall is low. As a rule of thumb, the method finds about half of the abuse. Therefore the results can be seen as an indicator of a more pervasive problem.

To compare this to the current state of the art, we refer to Wiegand \textit{et al}~\cite{wiegand2019detection}, who demonstrate that data-driven classification approaches leverage bias in the dataset to obtain an inflated result. The median F1 they find for a set of well-known systems, tested across domains to reduce this bias, is 0.617, showing that our performance is in keeping with the current state of the art. Furthermore our approach carries a much reduced risk of unwanted bias, such as more false positives for ethnic minorities or women~\cite{hardt2016equality,bolukbasi2016man,caliskan2017semantics}, that might reduce confidence in the findings presented here, since we don't use indiscriminate features.


\subsection*{Collecting Tweets}

The corpus was created by collecting tweets in real-time using Twitter's streaming API. We began immediately to collect any candidate who had been entered into Democracy Club's database\footnote{\small{\url{https://https://democracyclub.org.uk}}} who had Twitter accounts. We used the API to follow the accounts of all candidates over the period of interest. This means we collected all the tweets sent by each candidate, any replies to those tweets, and any retweets either made by the candidate or of the candidate's own tweets. Note that this approach does not collect all tweets which an individual would see in their timeline, as it does not include those in which they are just mentioned. We took this approach as the analysis results are more reliable  due to the fact that replies are directed at the politician who authored the tweet, and thus, any abusive language is more likely to be directed at them. Data were of a low enough volume not to be constrained by Twitter rate limits.

\subsection*{Ethics and Data Sharing}

Ethics approval was granted to collect the data through application 25371 at the University of Sheffield. All data used are in the public domain, and only public figures are identified by name in this work. Due to the sensitive nature of the data, it cannot be made public except in aggregate. See the supplementary materials~\footnote{\small{\url{https://gate-socmedia.group.shef.ac.uk/election-analysis-and-hate-speech/ge2019-supp-mat/}}} for aggregate data. An evolving process is in place to manage experimenter exposure to disturbing material. Readers are warned at the start of this work that they may find the language they encounter distressing.

\section*{Findings}

Table~\ref{tab:corpusstats} gives overall statistics of the corpus, which contains a total of 184,014 candidate-authored original tweets, 334,952 retweets and 131,292 replies. 3,541,769 replies to politicians were found, of which abuse was found in 4.46\%. The second row gives similar statistics for the 2017 general election period, allowing comparison to be made. It is evident that the level of abuse received by political candidates has risen in the intervening two and a half years.

\begin{table}[h!]
\caption{Number of tweets, retweets and replies by candidates are given, alongside replies received, of which abusive, and the percentage thereof. The first line shows findings for the six week election campaign period. The second line contrasts this with the corresponding period for the 2017 UK general election.}
\label{tab:corpusstats}
      \begin{tabular}{lllllll}
        \hline
        \textbf{Period} & \textbf{Original} & \textbf{MP} & \textbf{MP} & \textbf{Replies} & \textbf{Abusive repl-} & \textbf{\%}\\
         & \textbf{MP tweets} & \textbf{retweets} & \textbf{replies} & \textbf{to MPs} & \textbf{ies to MPs} & \textbf{Abusive} \\ \hline
        3 Nov--15 Dec 2019 & 184,014 & 334,952 & 131,292 & 3,541,769 & 157,844 & 4.46\\ \hline
        29 Apr--9 Jun 2017 & 126,216 & 245,518 & 71,598 & 961,413 & 31,454 & 3.27\\ \hline
      \end{tabular}
\end{table}

\begin{figure}[h!]
  \centering
\includegraphics[width=0.95\textwidth]{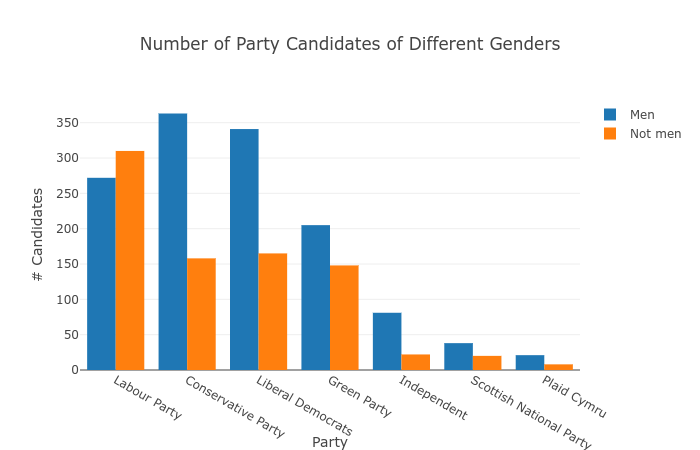}
  \caption{Gender representation of candidates for the major political parties.}
   \label{fig:cands-genders}
\end{figure}

The histogram in figure~\ref{fig:cands-genders} shows the gender balance of candidates representing each party. Four candidates identify as non-binary. These, along with women, including one transgender woman, form the ``not men'' category above. We can see that aside from in the Labour party, men are better represented.

\subsection*{Candidates}

Table~\ref{tab:mostabused} shows the ten most abused candidates across the period studied (November 3rd up to and including December 15th).

\begin{table}[h!]
\caption{Ten most abused candidates by volume.}
\label{tab:mostabused}
      \begin{tabular}{lllll}
        \hline
        \textbf{Name} & \textbf{Abusive replies} & \textbf{All replies} & \textbf{\% Abusive} & \textbf{\% of total}\\ \hline
        Boris Johnson & 34,256 & 565,396 & 6.06 & 21.70\\
        Jeremy Corbyn & 33,782 & 636,630 & 5.31 & 21.40\\
        Matthew Hancock & 12,156 & 186,543 & 6.52 & 7.70\\
        Michael Gove & 7,255 & 82,240 & 8.82 & 4.60\\
        David Lammy & 6,261 & 106,594 & 5.87 & 3.97\\
        Jo Swinson & 3,819 & 110,533 & 3.46 & 2.42\\
        James Cleverly & 3,571 & 58,856 & 6.07 & 2.26\\
        Jacob Rees-Mogg & 3,342 & 48,311 & 6.92 & 2.12\\
        Sajid Javid & 3,082 & 57,712 & 5.34 & 1.95\\
        Diane Abbott & 2,262 & 52,279 & 4.33 & 1.43\\ \hline
      \end{tabular}
\end{table}

\subsection*{Abuse Found}

In some cases it is possible to label the abuse as sexist, racist, religious, homophobic or political (``Tory scum'' for example), although in most cases abuse has no type (e.g. just ``idiot!''). Ethnic and religious minorities are underrepresented and therefore it is not possible to acquire reliable statistics for so short a time period comparing their experience to white candidates, but it is possible to see how men and women/non-gender-conforming candidates' experiences differ. Whilst prominent individuals may receive consistently high abuse levels amounting to as much as 6 or 7\% of their Twitter replies, on average male candidates received just over 1.2\% abuse, and women, 0.9\%. Men received almost twice as much abuse focused on their politics, as figure~\ref{fig:cands-genders} shows. Men received less than half as much sexist abuse. Men received more racist abuse.

\begin{figure}[h!]
  \centering
\includegraphics[width=0.95\textwidth]{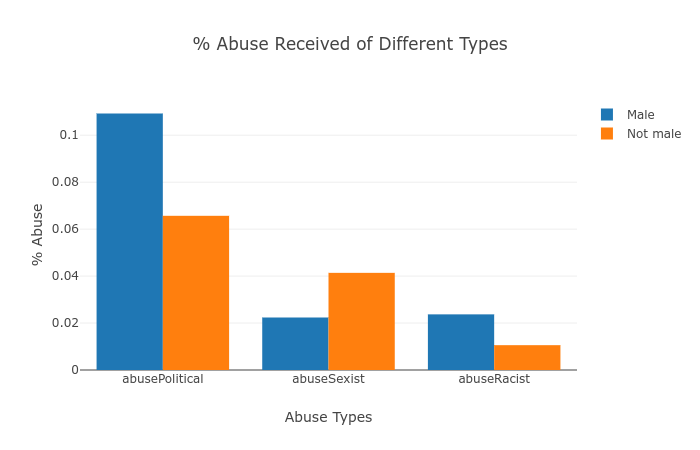}
  \caption{Political, sexist and racist abuse received by male vs. non-male candidates as a macro-averaged percentage of all replies received.}
   \label{fig:cands-genders}
\end{figure}

Table~\ref{tab:commonabuseterms} shows the most frequent abusive terms found across all types, followed by the most frequent politically abusive terms. Finally we see the most frequent sexist abusive terms. Inclusion of terms is heuristic; various sources have been combined, and further terms added through observation as the system has matured over several years. Yet there may be some terms we have overlooked despite our best efforts. Phrases that trip off the tongue may get to the top of these tables, whereas the long tail may contain more diverse ways of expressing a sentiment cluster that is harder for people to unite around words for. Below the table are word clouds, which include more terms than just the top ten. (If an abuse term of another type appears frequently with for example political abuse, it may appear in the word cloud.) Religious and homophobic abuse are too rare in the short time frame to produce interesting results (and confounded by much discussion of Boris Johnson's quote ``tank-topped bum boys''). Racism is in evidence but being rare, is better discussed in the context of a larger data sample, as in our previous work~\cite{gorrell2019race}, where ethnic minority politicians are found to receive more racist and sexist abuse.

\begin{table}[h!]
\caption{Most frequently found abusive terms for all abuse types and for political and sexist abuse, alongside count for the whole corpus.}
\label{tab:commonabuseterms}
      \begin{tabular}{ll|ll|ll}
        \hline
        \textbf{All abuse terms} & \textbf{Count} & \textbf{Political abuse terms} & \textbf{Count} & \textbf{Sexist abuse terms} & \textbf{Count}\\ \hline
        fuck off & 8,342 & tory scum & 284 & witch & 224\\
        idiot & 6,766 & remoaner & 276 & stupid woman & 149\\
        twat & 4,337 & fuck off commie & 113 & you stupid woman & 148\\
        coward & 2,649 & tory bullshit & 108 & stupid man & 108\\
        idiots & 2,632 & tory twat & 79 & you stupid man & 104\\
        scum & 2,324 & tory cunt & 60 & you silly man & 101\\
        cunt & 2,068 & tory bastards & 57 & silly man & 88\\
        moron & 1,979 & bloody tories & 54 & you silly woman & 78\\
        piss off & 1,866 & tory shit & 50 & stupid boy & 69\\
        wanker & 1,390 & fucking tory & 49 & silly boy & 65\\ \hline
      \end{tabular}
\end{table}

\begin{figure}[h!]
  \centering
\includegraphics[width=0.95\textwidth]{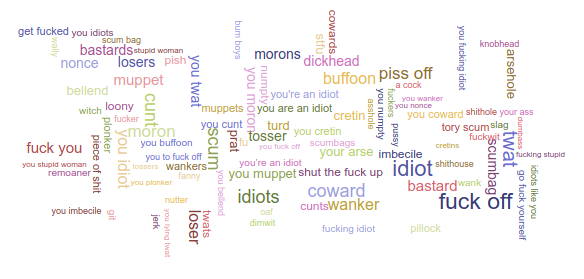}
  \caption{Word cloud for terms of all abuse types}
   \label{fig:wordcloudall}
\end{figure}

\begin{figure}[h!]
  \centering
\includegraphics[width=0.95\textwidth]{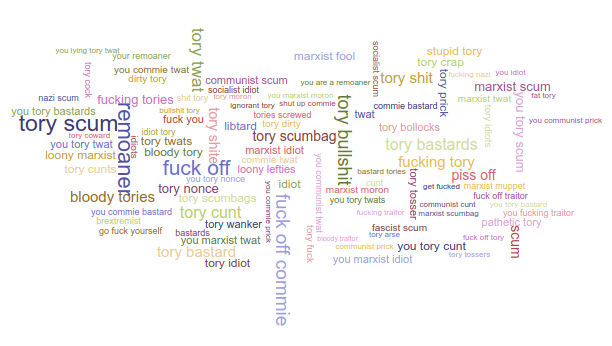}
  \caption{Word cloud for political abuse terms}
   \label{fig:wordcloudpolitical}
\end{figure}

\subsubsection*{Sexism}

Table~\ref{tab:mostabusedsexist} shows all individuals receiving more than 0.2\% sexist abuse in their replies and over 50 total in the period November 3rd to December 15th, ordered by total number of sexist abusive replies. Recall that accuracy is around 0.8, so smaller numbers of items detected might be considered less reliable.

\begin{table}[h!]
\caption{Individuals receiving more than 0.2\% sexist abuse and at least 50 sexist items in total}
\label{tab:mostabusedsexist}
      \begin{tabular}{llll}
        \hline
        \textbf{Name} & \textbf{Sexist abuse} & \textbf{All replies} & \textbf{\% Sexist}\\ \hline
        Jo Swinson & 464 & 110,533 & 0.42\\
        Diane Abbott & 202 & 52,279 & 0.39\\
        Caroline Lucas & 165 & 26,941 & 0.61\\
        Jess Phillips & 123 & 56,781 & 0.22\\
        Priti Patel & 122 & 39,616 & 0.31\\
        Anna Soubry & 102 & 30,912 & 0.33\\
        Yvette Cooper & 58 & 11,360 & 0.51\\
        Margaret Hodge & 50 & 4,915 & 1.01\\ \hline
      \end{tabular}
\end{table}

\begin{figure}[h!]
  \centering
\includegraphics[width=0.95\textwidth]{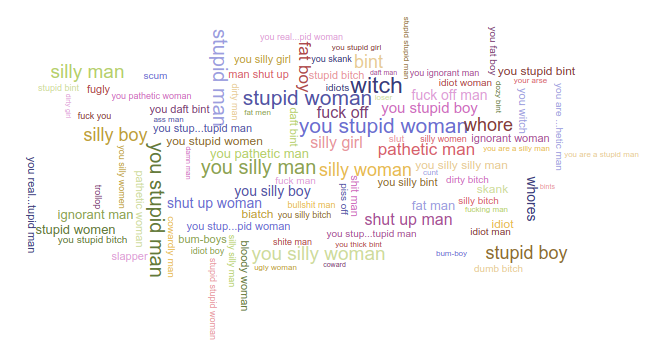}
  \caption{Word cloud for sexist abuse terms.}
   \label{fig:wordcloudsexist}
\end{figure}

We can see from the word cloud in figure~\ref{fig:wordcloudsexist} that sexist abuse toward men was counted and did occur in the corpus, though specifically misandristic terms are not readily available for men (equivalents for ``witch'', ``bint'' etc.) and choice of terms is somewhat subjective. Excluding those with fewer than 0.2\% sexist replies and fewer than 50 sexist replies overall, the candidates receiving the highest percentage of sexist abuse are given in table~\ref{tab:mostabusedsexist}, and are all women. Jo Swinson received the most sexist abuse of any candidate in this period, although by volume, Boris Johnson was not far behind with 351 items; for him, however, that only constituted 0.08\% of all replies received.

\subsection*{Abuse Fluctuation over Time}

\begin{table}[h!]
\caption{Tweet statistics on a per-week basis for the six week campaign period}
\label{tab:corpusstatsperweek}
      \begin{tabular}{lllllll}
        \hline
        \textbf{Period} & \textbf{Original} & \textbf{MP retweets} & \textbf{MP replies} & \textbf{Replies} & \textbf{Abusive repl-} & \% \textbf{Abusive}\\
        & \textbf{MP tweets} & & & \textbf{to MPs} & \textbf{ies to MPs} & \\ \hline
        Nov 3--Nov 9 & 18,633 & 40,683 & 14,456 & 464,473 & 17,854 & 3.84\\
        Nov 10--Nov 16 & 19,845 & 40,110 & 14,651 & 444,045 & 20,742 & 4.67\\
        Nov 17--Nov 23 & 30,445 & 57,764 & 19,372 & 547,748 & 22,007 & 4.02\\
        Nov 24--Nov 30 & 35,254 & 62,688 & 23,674 & 572,976 & 27,666 & 4.83\\
        Dec 1--Dec 7 & 37,615 & 65,601 & 24,237 & 590,781 & 28,151 & 4.77\\
        Dec 8--Dec 14 & 42,222 & 78,106 & 34,902 & 921,746 & 41,421 & 4.49\\ \hline
      \end{tabular}
\end{table}

On a week by week basis, we see a rising level of abuse toward candidates, as shown in table~\ref{tab:corpusstatsperweek}. The graph below shows that for the majority of the period, this was due to rising abuse toward Conservative candidates, which was not echoed in responses to either Labour or Liberal Democrat candidates.\footnote{Party disparity not due to greater engagement on Twitter by Conservative candidates. Tweets authored:
Wk 1, Cons 11,311, Lab 19,965;
Wk 2, Cons 10,921, Lab 19,648;
Wk 3, Cons 15,404, Lab 25,973;
Wk 4, Cons 19,065, Lab 33,468.} Abuse toward Labour candidates however rose sharply following their decisive defeat.

\begin{figure}[h!]
  \centering
\includegraphics[width=0.95\textwidth]{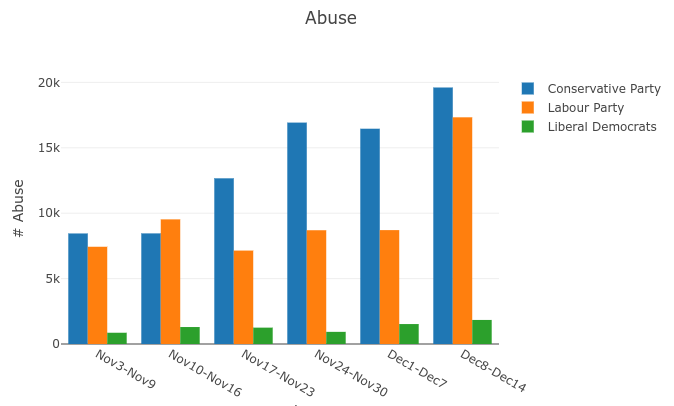}
  \caption{Abuse received by volume for candidates of the three biggest parties on a per week basis.}
   \label{fig:perweekperparty}
\end{figure}

\begin{figure}[h!]
  \centering
\includegraphics[width=0.95\textwidth]{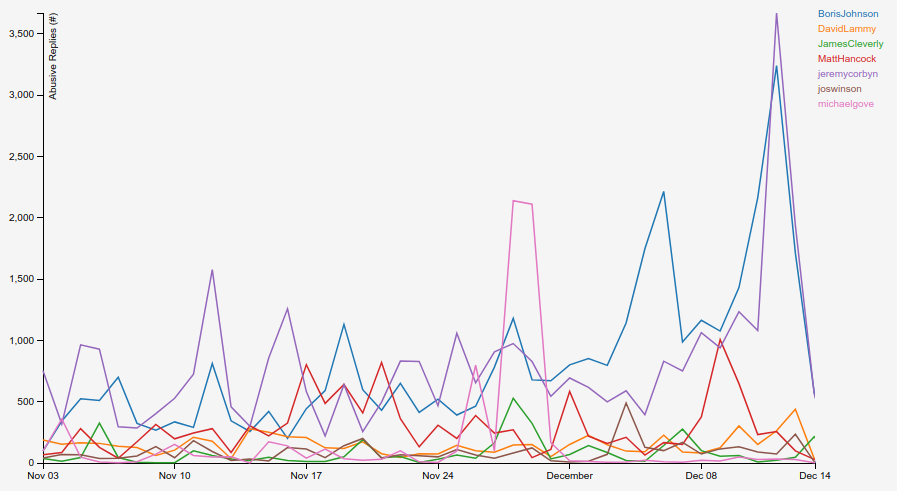}
  \caption{Abuse volume timelines for the seven candidates who received the most abuse in total.}
   \label{fig:mostabusedtimeline}
\end{figure}

In figure~\ref{fig:mostabusedtimeline} we see the timeline up to and including December 14th for the seven candidates who received the most abuse by volume. The two main party leaders, Jeremy Corbyn and Boris Johnson, received the most abuse by far. There is somewhat of an increase across the period in abuse toward Mr Johnson. Furthermore, prominent Conservatives also receive significant levels of abuse. Mr Gove receives a prominent spike around the time of the climate debate. Mr Johnson receives the highest spike before the election at the time of the BBC Prime Ministerial Debate on December 6th, echoing a pattern discussed below where television appearances lead to a spike in Twitter abuse toward Mr Johnson but less so toward Mr Corbyn. On a per-day basis, the highest abuse spike of the period is in that received by Mr Corbyn, however, at the time of the election.

\begin{figure}[h!]
  \centering
\includegraphics[width=0.95\textwidth]{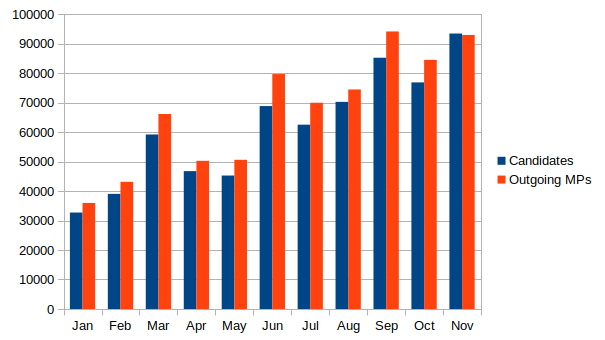}
  \caption{Abuse received by candidates alongside outgoing MPs per month across 2019. Individuals who are both previous MPs and candidates appear in both columns.}
   \label{fig:abusepermonth2019}
\end{figure}

In figure~\ref{fig:abusepermonth2019} we show abuse received per month by individuals who are running in 2019, alongside that received by individuals who were MPs in the previous parliament.\footnote{The chart of abusive tweet count per month since January naturally doesn't contain all candidates, since these were only announced in November. It is based on our previous data collection, and contains only candidates that also stood in 2017. However, considering most abuse/replies are received by the most prominent individuals, the overlap is very high. 99\% of abuse in the new dataset also appears in the old one (and 95\% of replies). For the calendar month of November, in the new dataset 94,566 abusive replies to candidates were found whereas the graph shows 93,516.} Outgoing MPs who are also candidates appear in both columns. Most abuse is received by a handful of prominent individuals that appear in both categories. The graph shows that abuse toward politicians by volume has risen steeply across the year. As a percentage of replies received, there has been an increase of around 1\% as calculated on outgoing MPs. (In November, candidates who aren't previous MPs are likely to have stepped up their engagement level, and resigning MPs the opposite, making it the only month where candidates receive more abuse. In the previous months, we are counting abuse for individuals that are only ``candidates'' in hindsight--they didn't actually announce their candidacy until November.)

As mentioned at the beginning of the section, in the same six week period surrounding the 2017 election candidates drew 31,454 abusive tweets, compared with 157,844 this time, which is less than 20\%, and as a percentage of replies drew 3.3\% compared with 4.5\% this time, suggesting an upward trend in the abuse politicians are exposed to that was also evident across 2019 and across the campaign period itself.

\subsubsection*{Experience of standing and resigning MPs since January}

Twelve Conservative or formerly Conservative MPs stated opposition to the party's Brexit policy as the precipitating factor for their standing down. Three Labour or former Labour MPs cite concerns about the climate or leadership of the Labour Party. Additional to this, further MPs standing down, such as Louise Ellman, have had rocky relationships with their party, which affected their decision to stand again.\footnote{https://www.theguardian.com/politics/2019/oct/31/which-mps-are-standing-down-at-the-2019-general-election} Their rebel status might be a contributing factor to the abuse they received. A graphic showing changes in MPs' party membership can be viewed here, with green stars indicating those who chose not to stand again: \url{https://gate4ugc.blogspot.com/2019/11/which-mps-changed-party-affiliation.html}

However several have also explicitly referred to abuse as the reason or one of their reasons for standing down: Nicky Morgan,\footnote{https://twitter.com/NickyMorgan01/status/1189625485124354048} Caroline Spelman,\footnote{https://www.bbc.co.uk/news/uk-england-birmingham-49593559} Teresa Pearce,\footnote{https://www.stylist.co.uk/long-reads/women-mps-standing-down-uk-online-abuse-election/325744} Heidi Allen,\footnote{https://www.theguardian.com/politics/2019/oct/29/lib-dem-mp-heidi-allen-stand-down-next-general-election} Mark Lancaster.\footnote{https://www.bbc.co.uk/news/uk-england-beds-bucks-herts-50275204}

\begin{figure}[h!]
  \centering
\includegraphics[width=0.95\textwidth]{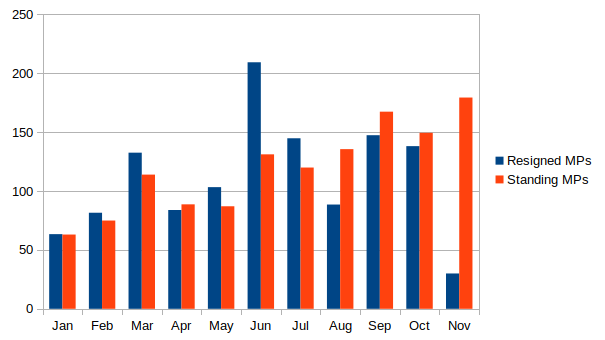}
  \caption{Macro-averaged abuse per month per individual for those MPs who chose to stand again vs those who chose not to.}
   \label{fig:resigners}
\end{figure}

In figure~\ref{fig:resigners} we directly compare average abuse per month received by MPs who chose not to stand again against those who did choose to stand again. We see that in all bar one of the earlier months of the year those individuals received more abuse, and particularly in June. When considered as a percentage of replies received, the MPs that stood down had on average\footnote{Macro-average: the percentage is calculated per individual and then averaged, to avoid prominent individuals dominating the overall result} more abuse than the ones that are standing again in every single month of the year.


\subsection*{Topics}

\begin{figure}
\centering
\begin{minipage}{.65\textwidth}
  \includegraphics[width=\linewidth]{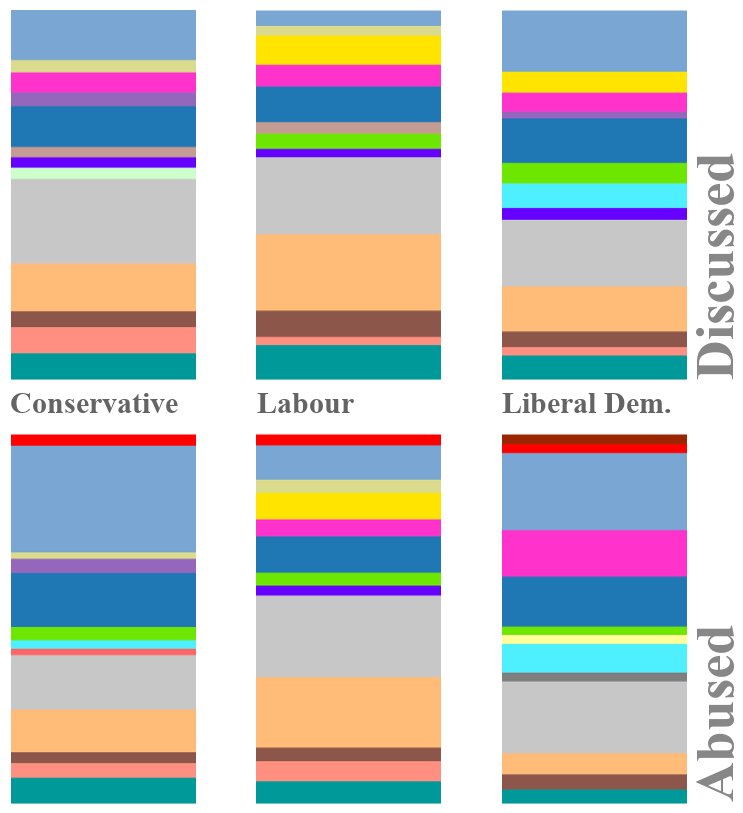}
\end{minipage}%
\begin{minipage}{.35\textwidth}
  \centering
  \includegraphics[width=0.9\linewidth]{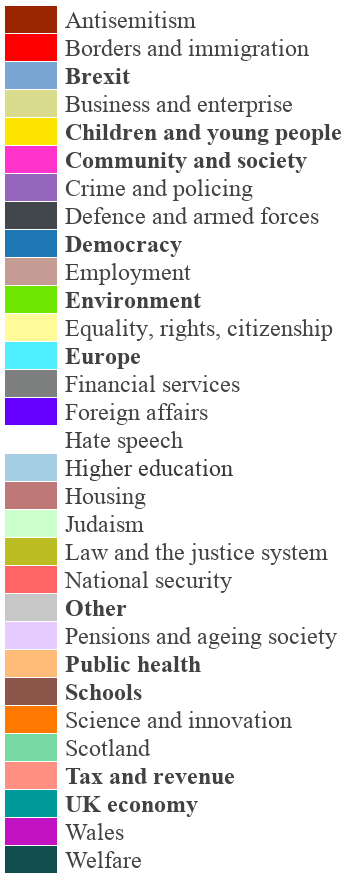}
\end{minipage}
 \caption{Twelve topics most discussed by candidates of each party over the six week period, above the topics that drew the most abuse.}
\label{fig:topicssummary}
\end{figure}

The top twelve topics mentioned by candidates in their tweets for each party are shown in figure~\ref{fig:topicssummary}, with the remainder in the ``other'' category (long tail). Important, recurrent subjects are highlighted in bold type in the key. We see the overall topic balance in tweets by candidates of the Conservatives, Labour and Liberal Democrats. We see that public health features prominently for all parties, but that Brexit occupies diverse positions, with Labour de-prioritising this subject and the Liberal Democrats fore-fronting it. The prominence of the topic of democracy is to be expected at election time. The Conservatives give airtime to taxation where Labour favour the economy and children/young people. Only the Liberal Democrats give significant airtime to Europe. The Conservatives don't focus on the environment. All parties discuss community and society (religious groups) and schools.

Figure~\ref{fig:topicssummary} in the lower part shows what topics attracted abusive responses when discussed by candidates, and shows that not all topics attract these responses equally. The Conservative party receive disproportionate abuse when they talk about Brexit. Tax and revenue is a safe topic for them, as is business and enterprise. Borders and immigration, Europe and policing all draw fire for the Conservatives.

Borders and immigration is a topic that also draws fire for the Labour Party, as does Brexit and democracy. Public health is a safe topic for Labour. Business and enterprise draws fire. For the Liberal Democrats, Brexit, democracy and Europe are brave topics, receiving disproportionate abuse, as is community and society. The environment is a safe topic for them, as indeed it is for Labour.

We tend to see a greater diversity of topics among those that drew abusive replies than in the topics candidates chose to talk about. This is partly because inflammatory topics are ones that candidates aren't focusing on to a great extent and partly because abuse is distributed very unevenly, clumping mainly on a handful of the tweets, so can potentially make much of a topic that was little discussed by the candidates.

\subsubsection*{Topics: Conservatives}

\begin{figure}
\centering
\begin{minipage}{.65\textwidth}
  \includegraphics[width=\linewidth]{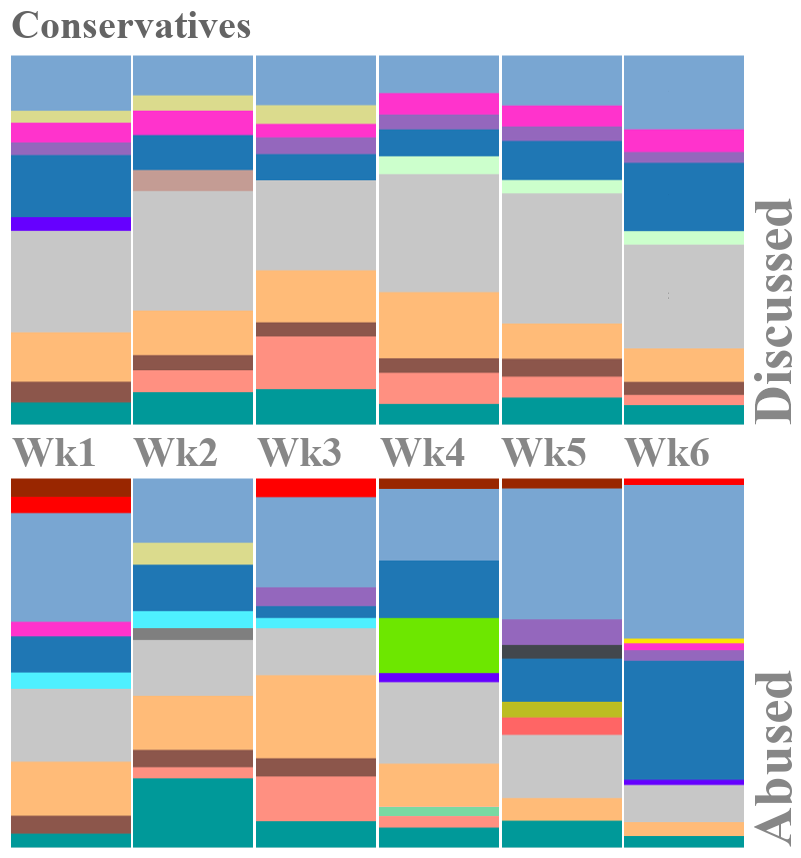}
\end{minipage}%
\begin{minipage}{.35\textwidth}
  \centering
  \includegraphics[width=0.9\linewidth]{key.png}
\end{minipage}
 \caption{Twelve topics most discussed by Conservative candidates per week, above the topics that drew the most abuse.}
\label{fig:topicsperweekcons}
\end{figure}

As the weeks progress, we see in figure~\ref{fig:topicsperweekcons} that Conservative candidates have somewhat reduced their focus on Brexit, introducing public health as their dominant topic and increasing focus on taxation. This reflects the topics on which they are popular (at least on Twitter, and may or may not reflect an evolution of focus outside of Twitter). In terms of the topics that drew abusive replies, we see increases in week 2 for the economy and in week 4, around the time of the climate debate, for the environment, as shown in the lower part of the figure. Brexit has been a topic that has consistently drawn abuse for the Conservatives. In week 5, focus returns to Brexit.

\subsubsection*{Topics: Labour Party}

\begin{figure}
\centering
\begin{minipage}{.65\textwidth}
  \includegraphics[width=\linewidth]{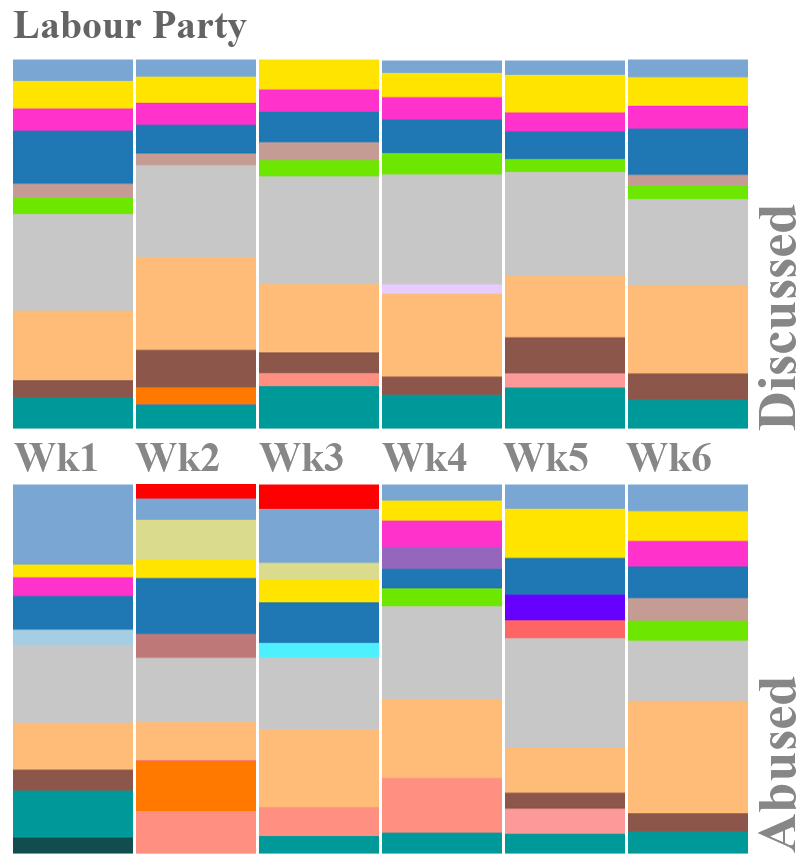}
\end{minipage}%
\begin{minipage}{.35\textwidth}
  \centering
  \includegraphics[width=0.9\linewidth]{key.png}
\end{minipage}
 \caption{Twelve topics most discussed by Labour candidates per week, above the topics that drew the most abuse.}
\label{fig:topicsperweeklab}
\end{figure}

Labour candidates have retained their focus on public health, as shown in figure~\ref{fig:topicsperweeklab}, which is a popular topic for them, and increased their focus on the economy. Earlier in November, they received abuse particularly for tweets about Brexit, as shown in the lower part of the figure, but this did not continue.

\subsubsection*{Topics: Liberal Democrats}

\begin{figure}
\centering
\begin{minipage}{.65\textwidth}
  \includegraphics[width=\linewidth]{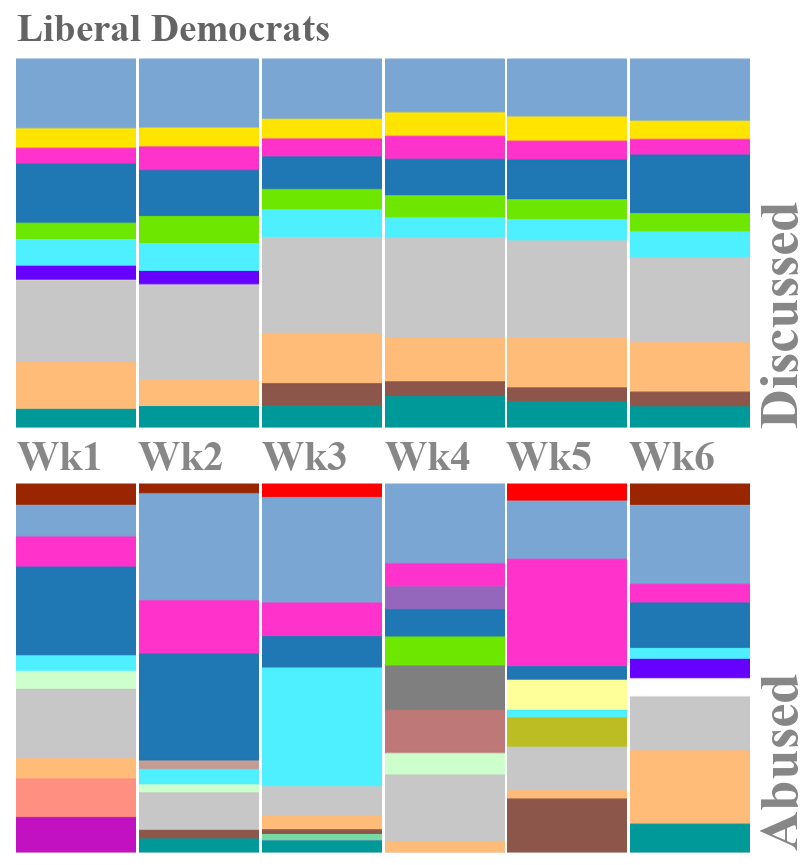}
\end{minipage}%
\begin{minipage}{.35\textwidth}
  \centering
  \includegraphics[width=0.9\linewidth]{key.png}
\end{minipage}
 \caption{Twelve topics most discussed by Liberal Democrat candidates per week, above the topics that drew the most abuse.}
\label{fig:topicsperweeklibd}
\end{figure}

As shown in figure~\ref{fig:topicsperweeklibd}, the Liberal Democrats have been true to their focus on Brexit despite receiving disproportionate abuse on this topic. Of the three parties, they are the one that has shown the most consistent interest in the environment. In terms of the topics drawing abusive replies, democracy was prominent earlier in November, with Europe and Brexit also highly prominent across the month. Liberal Democrat candidates receive visibly more abuse than candidates of other parties when they tweet about community and society (religious and gender identities), as shown in the lower part of the figure. This appears to arise from their being outspoken regarding minority rights.

\subsection*{Debate Periods and Significant Incidents}

We now give per-hour timelines for key events in the campaign period. Per-hour timelines can give more information about the way Twitter users responded to events.

\subsubsection*{November 19th ITV Prime Ministerial Debate}


Abuse toward Boris Johnson spiked around the time of the November 19th ITV debate, to his highest level since the start of November. Abuse did not especially spike toward Jeremy Corbyn around the time of the debate, though generally speaking Jeremy Corbyn draws more abuse on Twitter than Boris Johnson, and three times in November already had spikes that exceeded that of Mr Johnson at the time of this debate.

Abuse levels as a percentage toward Mr Johnson and Mr Corbyn were lower around the time of the debate, likely because the debate inspired additional non-abusive Twitter users to engage with them. However abuse levels toward Jo Swinson and Theresa May as a percentage were higher around this time as their comments about the debate inspired uncivil responses.

\begin{figure}[h!]
  \centering
\includegraphics[width=0.95\textwidth]{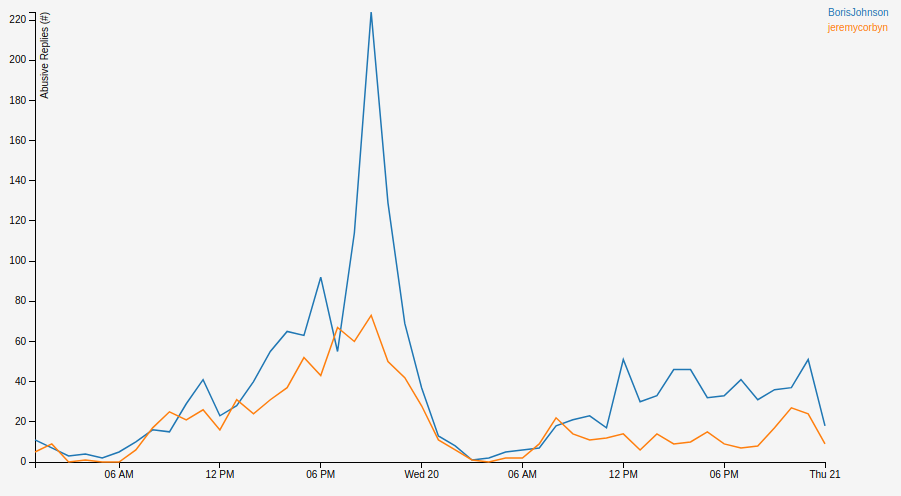}
  \caption{Per-hour timeline for the two participants covering the two day period around the November 19th ITV prime ministerial debate.}
   \label{fig:nov19}
\end{figure}

Figure~\ref{fig:nov19} gives an hour-by-hour timeline of the two party leaders for the two day window spanning the debate (Monday midnight until Wednesday midnight).

\subsubsection*{November 22nd BBC Question Time Leaders’ Special}


At the the time of airing of the Question Time programme, Boris Johnson drew the highest peak of abuse on Twitter in absolute terms, echoing what we saw on November 19th though to a less pronounced degree, despite the fact that Mr Corbyn usually draws more abuse on Twitter.

As a percentage, the abuse received by both men across November 22nd-23rd was more similar to background, unlike the 19th where Mr Corbyn's abuse level was substantially lower and Mr Johnsons, higher.

Jo Swinson received a bigger increase in replies proportionally compared with background than Mr Johnson or Mr Corbyn for her first appearance in an event of this kind on the 22nd, and she received less abuse around that date, both in absolute terms and substantially so as a percentage.

In summary, responses to television appearances for Jo Swinson and Jeremy Corbyn appeared to be more civil.

\begin{figure}[h!]
  \centering
\includegraphics[width=0.95\textwidth]{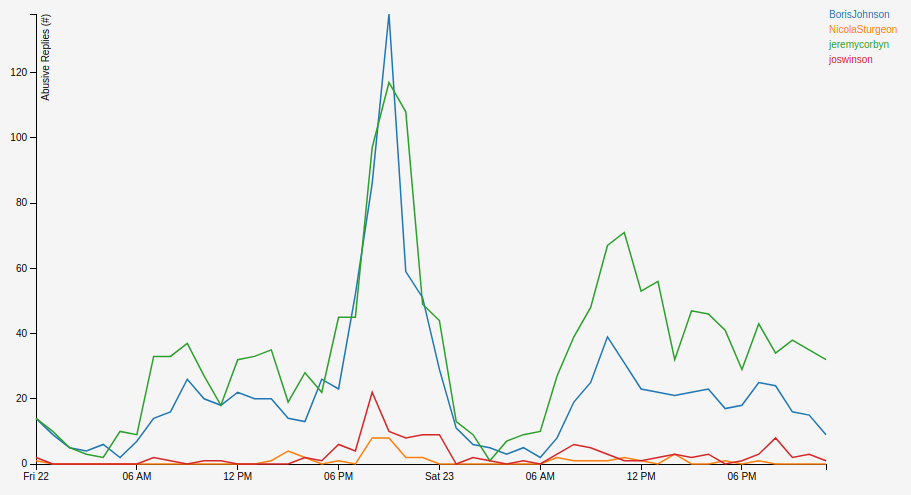}
  \caption{Per-hour timeline for participants covering the two day period around the November 22nd BBC Question Time leaders' special.}
   \label{fig:nov22}
\end{figure}

The hourly timeline in figure~\ref{fig:nov22} shows that again, Boris Johnson received more abuse on the night. Jo Swinson received much less abuse. Ms Swinson is one of the most abused candidates on Twitter, being the sixth most abused candidate according to our data from the start of the month (see below), but due to the Zipfian nature of the abuse distribution, this still amounts to a fraction of that of Mr Corbyn or Mr Johnson. Nicola Sturgeon received the least abuse of all the party leaders represented in the programme.

In comparison with November 19th, less abuse was directed at Mr Johnson and Mr Corbyn on the 22nd in absolute terms, and the differential between the two men was less.

\subsubsection*{November 28th Channel 4 Climate Debate, November 29th BBC Election Debate and London Bridge stabbing}



\begin{figure}[h!]
  \centering
\includegraphics[width=0.95\textwidth]{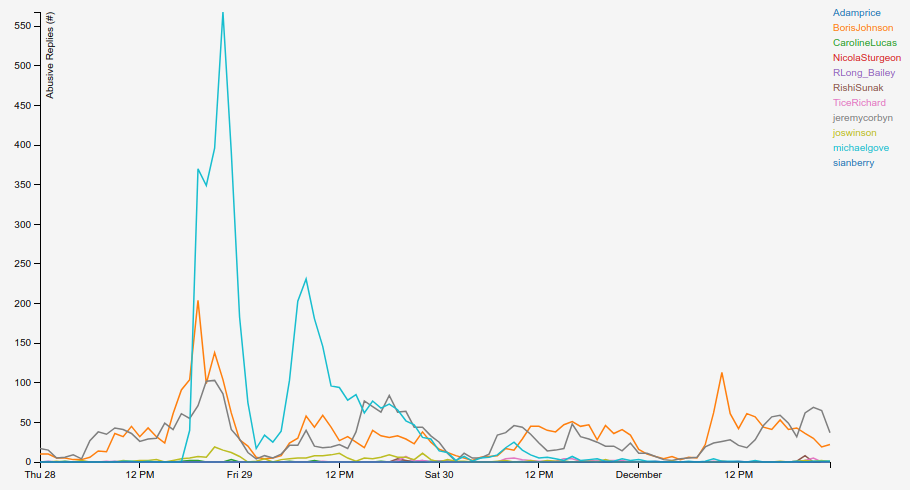}
  \caption{Per-hour timeline for participants covering the period around the Channel 4 climate debate and BBC election debate.}
   \label{fig:nov28dec1}
\end{figure}

The period of November 28th and 29th was eventful, featuring two election-related television events and also encompassing an Islamist attack in which two people died. The main event drawing fire on Twitter was Michael Gove's attempt to participate in Boris Johnson's place on Thursday, and his subsequent attitude about being refused. Both Mr Gove and Mr Johnson drew more abuse on Twitter on Thursday night than any of the actual participants. Participants in both events did not particularly come under fire. Mr Corbyn received somewhat of a peak of abuse during each event, and Ms Swinson received some abuse, but other participants received very little abuse. The timeline in figure~\ref{fig:nov28dec1} shows a large peak in abuse toward Michael Gove around the time of the Thursday night climate debate.

\begin{figure}[h!]
  \centering
\includegraphics[width=0.95\textwidth]{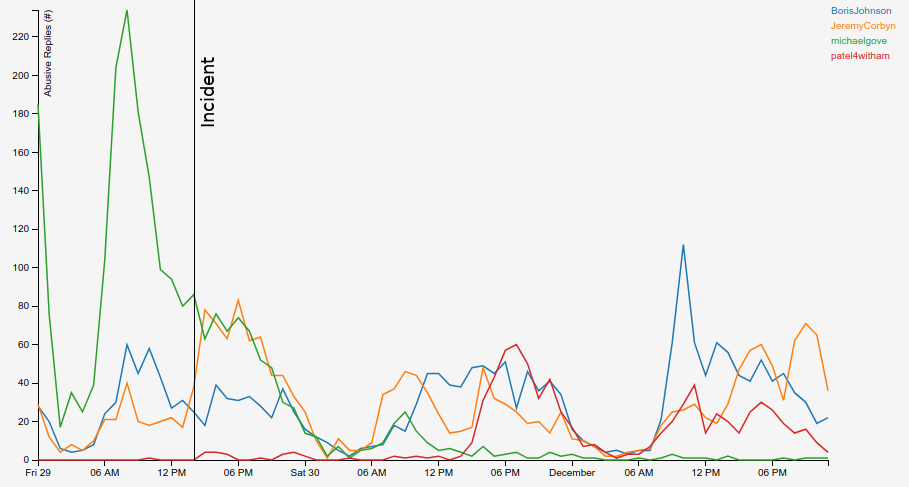}
  \caption{Per-hour timeline for the most abused individuals around the London Bridge stabbing. The time of the incident is highlighted using a vertical line.}
   \label{fig:londonbridge}
\end{figure}

Figure~\ref{fig:londonbridge} shows individuals receiving the most abuse around the time of the London Bridge stabbing, in a timeline on which the incident is annotated. Two tweets by Home Secretary Priti Patel drew more abuse to her than she usually receives.\footnote{\small{\url{https://twitter.com/patel4witham/status/1200792218749022213} \url{https://twitter.com/patel4witham/status/1200815822853328898}}} In the tweets she blames Labour government legislation for the release of the attacker. The abuse she received is predominantly of a general nature, but political (usually ``tory \_\_\_'') and sexist (around half of which is ``witch'') types appear. Racist or religious abuse toward Ms Patel is not in evidence.

\subsubsection*{December 6th BBC Prime Ministerial Debate}


\begin{figure}[h!]
  \centering
\includegraphics[width=0.95\textwidth]{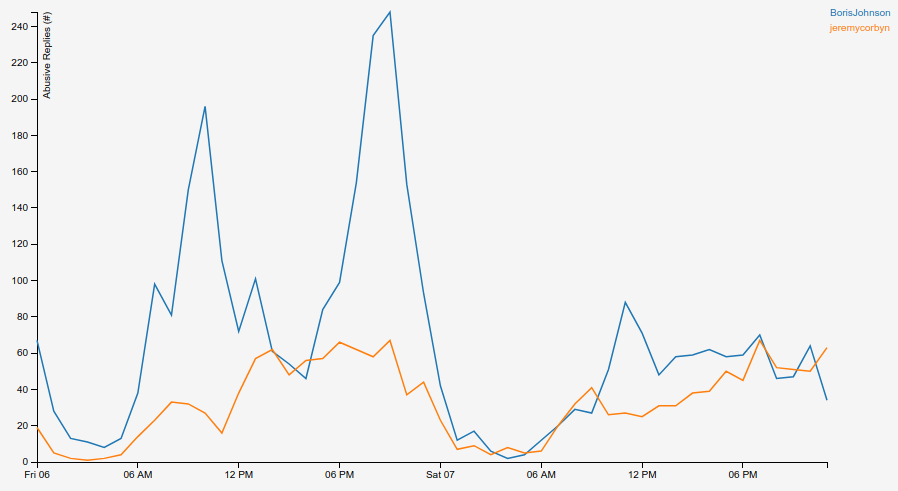}
  \caption{Per-hour timeline for participants covering the period around the BBC prime ministerial debate.}
   \label{fig:dec6}
\end{figure}

Figure~\ref{fig:dec6} shows the per-hour timeline for the December 6-7 period, spanning the BBC prime ministerial debate on the evening of the 6th. Mr Johnson's peak during the debate at 8.30pm on Friday evening echoes earlier data where television events result in a peak of abuse for Mr Johnson, more-so than for Mr Corbyn. The peak on Friday morning follows Mr Johnson's late Thursday tweet\footnote{https://twitter.com/BorisJohnson/status/1202543318133092352} of a somewhat controversial ``selfie'' from his Thursday morning appearance with Holly Willoughby and Phillip Schofield on ``This Morning''.

\subsubsection*{Election Period}

\begin{figure}[h!]
  \centering
\includegraphics[width=0.95\textwidth]{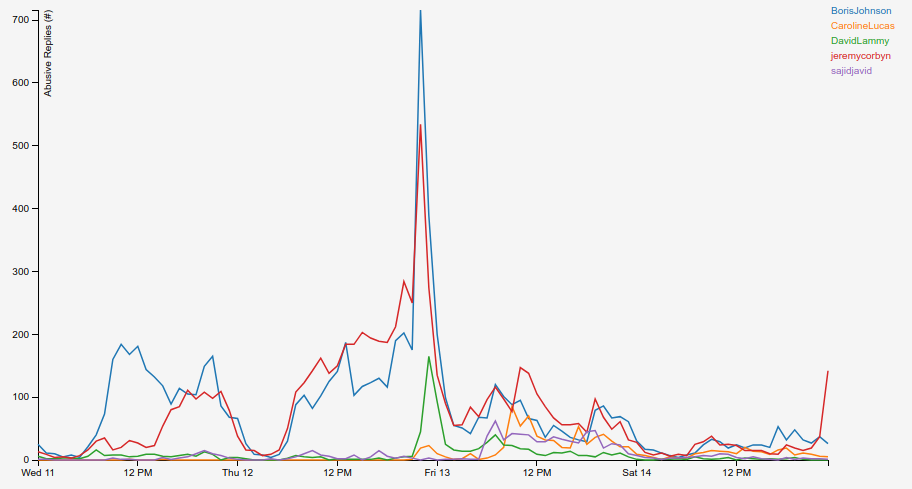}
  \caption{Per-hour timeline for the individuals receiving the most abuse in the two-day period covering the election on December 12th.}
   \label{fig:dec12}
\end{figure}

At the time of the election itself, the highest spike of abuse was received by Boris Johnson, echoing the pattern seen around television events, as shown in figure~\ref{fig:dec12}. However Jeremy Corbyn also received a high spike of abuse, as well as more abuse in the surrounding days. It is unusual to see Caroline Lucas and Sajid Javid among those who received the most abuse.\footnote{\small{Here are tweets that inspired a particularly abusive response around that period (in addition to several tweets from Boris Johnson and Jeremy Corbyn, that are unsurprisingly high profile): \url{https://twitter.com/CarolineLucas/status/1205403440144429056} \url{https://twitter.com/sajidjavid/status/1205371247040909313} \url{https://twitter.com/DavidLammy/status/1205258582029225984}}}

\section*{Conclusion and Future Work}

Between Nov 3rd and December 15th, we found 157,844 abusive replies to candidates' tweets (4.44\% of all replies received)--a low estimate of probably around half of the actual abusive tweets. Overall, abuse levels climbed week on week in November and early December, as the election campaign progressed, from 17,854 in the first week to 41,421 in the week of the December 12th election. The escalation in abuse was toward Conservative candidates specifically, with abuse levels towards candidates from the other two main parties remaining stable week on week; however, after Labour's decisive defeat, their candidates were subjected to a spike in abuse. Abuse levels are not constant; abuse is triggered by external events (e.g. leadership debates) or controversial tweets by the candidates. Abuse levels have also been approximately climbing month on month over the year, and in November were more than double by volume compared with January.

In terms of representation in the sample of election candidates with Twitter acccounts, gender balance is skewed heavily in favour of men for the Conservatives and LibDems; Labour in contrast has more female/non-binary than male candidates in our sample. Most abuse is aimed at Jeremy Corbyn and Boris Johnson, with Matthew Hancock, Jacob Rees-Mogg, Jo Swinson, Michael Gove, David Lammy and James Cleverly also receiving substantial abuse. Michael Gove received a great deal of personal abuse following the climate debate. Jo Swinson received the most sexist abuse.

The topic of Brexit draws abuse for all three parties. Conservative candidates initially move away from this, toward their safer topic of taxation, before returning to Brexit. Liberal Democrats continue to focus on Brexit despite receiving abuse. Labour candidates consistently don't focus on Brexit; public health is a safe topic for Labour. MPs who stood down received more abuse than those who chose to stand again in all but one month in the first half of 2019, and in June they received over 50\% more abuse.


\begin{backmatter}

\section*{Competing interests}

The authors declare that they have no competing interests.

\section*{Authors' contributions}

Genevieve Gorrell designed and performed the statistical analysis, developed the abuse classifier and wrote the greater part of the text. Mehmet Bakir wrote the software to extract summary data from indexed tweets. Ian Roberts managed the indexing of the raw tweets into our search engine. Mark Greenwood helped to produce some of the visualisations. Kalina Bontcheva provided direction and guidance, and wrote some of the text.

\section*{Acknowledgements}

This work was partially supported by the European Union under grant agreements No. 825297 ``WeVerify'' and No. 654024 ``SoBigData''.


\bibliographystyle{bmc-mathphys} 
\bibliography{ge2019-epj}      

\end{backmatter}
\end{document}